\documentclass[a4paper,12pt]{article}
\usepackage{color}
\usepackage{graphicx, color}
\usepackage{amsmath}
\usepackage{float}
\usepackage{tikz}
\usepackage[utf8]{inputenc}
\usepackage{authblk}
\usepackage[hidelinks]{hyperref}
\usepackage{titlesec}
\usepackage{lipsum}
\usepackage{titlesec}
\usepackage{lipsum}
\usepackage[top=90pt,bottom=90pt,left=80pt,right=78pt]{geometry}
\titleformat{\section}
{\normalfont\large\bfseries}{\thesection}{1em}{}

\title{\bf \Large Uniqueness from locality and BCFW shifts}
\author{Laurentiu Rodina}
\affil{Department of Physics, Princeton University, Princeton, NJ 08540}
\date{}

\numberwithin{equation}{section}
\makeatletter
\newcommand{\eqnum}{\leavevmode\hfill\refstepcounter{equation}\textup{\tagform@{\theequation}}}
\makeatother

\begin{document}
\maketitle
\begin{abstract}
We introduce a BCFW shift that can be used to recursively build the full Yang-Mills tree-level amplitude as a function of polarization vectors. Furthermore, in line with the recent results of \cite{Nima}, we conjecture that the Yang-Mills tree-level scattering amplitude is uniquely fixed by locality and demanding the usual asymptotic behavior under a sufficient number of shifts. Unitarity therefore emerges from locality and constructability. We prove this statement at the leading order in the soft expansion.
\end{abstract}
\section{ Introduction}

The traditional formulation of Quantum Field Theory is based on Feynman diagrams, which ensure that locality and unitarity are manifest at all times. But to accomplish this, Feynman diagrams introduce a large amount of unphysical redundancy, which hides the ultimate simplicity of scattering amplitudes in many theories \cite{simplest}. Nowhere is this more striking than in Yang-Mills and General Relativity, where gauge and diffeomorphism invariance lead to very complicated Feynman diagram expansions, containing thousands of terms even for five particle scattering. But quite surprisingly, with the right variables, such expressions can ultimately be collapsed into remarkably simple answers, such as the Parke-Taylor formula for Yang-Mills \cite{Parke}, or the Kawai-Lewellen-Tye relations for gravity \cite{klt}.

The S-matrix program on the other hand aims to replace the Lagrangian formulation by directly imposing physical principles on scattering amplitudes. Motivated by this hidden simplicity and dealing away with gauge invariance, the on-shell perspective recently lead to many computational and conceptual advances, chiefly through the use of recursion relations \cite{BCF,BCFW}. Other recent developments have revealed more and more previously unknown facets of scattering amplitudes: the twistor string picture \cite{twistor}, the BCJ duality \cite{bcj}, scattering equations \cite{CHY}, and many others.  However, in this approach the physical nature of scattering amplitudes - that is, particles scattering off of each other in local quantum interactions -  is completely lost in favor of more abstract properties and symmetries. The amplituhedron \cite{amplituhedron} is a prime example of this newer perspective. There, scattering amplitudes can be understood as volumes of certain polytopes, with locality and unitarity emerging from the geometry itself. 

The on-shell and off-shell techniques therefore capture very different aspects of scattering amplitudes, and the goal of this paper is to explore the interface of these two otherwise orthogonal perspectives. 

In section \ref{sec2}, we bridge the gap between on-shell recursions and Feynman amplitudes, by introducing a BCFW shift compatible with arbitrary polarization vectors. This shift is in fact the manifestly covariant form of the shift used in Ref. \cite{nima0}, and can be used in the usual way to recursively build the full tree-level amplitude, valid in any dimension and for all helicity configurations.

In section \ref{sec3}, we conjecture that the Yang-Mills tree-level amplitude is completely fixed by imposing locality (singularity structure given by propagators associated to cubic diagrams) and constructability (vanishing of poles at infinity under BCFW shifts).  Unitarity (factorization) is never used, but instead emerges as a consequence of uniqueness. This result is very similar to (and was motivated by), the recent results in \cite{Nima}, which showed that gauge invariance uniquely fixes the Yang-Mills and gravity amplitudes. It is worth noting that the needed asymptotic behaviors which will show up are more general than those required in the usual BCFW recursion, and in fact  we will not use the Cauchy theorem at any point. Instead of building the scattering amplitude directly via on-shell recursions (which would mean assuming unitarity), we simply argue that there is a unique local object satisfying the vanishing conditions for $z\rightarrow \infty$.

The strategy is almost identical to the one in \cite{Nima}: we show uniqueness order by order in the soft expansion by using induction. However, checking BCFW behavior is more complicated than checking gauge invariance. Instead of simply imposing vanishing under $e_i\rightarrow p_i$ for $n-1$ particles, now we must require some specific $\mathcal{O}(z^m)$ behavior under $ n(n-1)$ BCFW shifts $[i,j\rangle$ which involve $e_i$, $p_i$, $e_j$, and $p_j$. This makes finding a precise inductive argument more difficult. Even fixing only the leading term, which was immediate with gauge invariance, is a lot more involved. This time, we cannot even impose any shift involving the soft particle, as it interferes with momentum conservation. In fact, in this case the process is reversed: the lower point amplitude is fixed first, and the soft factor is fixed last. For this reason, we limit our discussion to the leading order, and conjecture that the same argument can be used for the subleading orders as well. Nevertheless, explicit checks of the all-order statement have been made up to five points.

\section{BCFW  with polarization vectors}\label{sec2}
We define our $[i,j\rangle$ shift in the following way:
\begin{align}
\begin{split}
\label{shift}
e_i &\rightarrow \hat{e}_i\, ,\\
e_j &\rightarrow \hat{e}_j+zp_i\frac{\hat{e}_i.\hat{e}_j}{p_i.p_j}\, ,\\
p_i&\rightarrow p_i+z \hat{e}_i\, ,\\
p_j&\rightarrow p_j-z \hat{e}_i\, ,
\end{split}
\end{align}
where $\hat{e}_i=e_i-p_i\frac{e_i.p_j}{p_i.p_j}$, and similarly for $\hat{e}_j$. The motivation for this peculiar shift, which generalizes the shift in Ref. \cite{nima0}, is that it maintains the on shell conditions $e_i.p_i=0$ and $e_j.p_j=0$. Alternativey, a simpler version of the shifts may used, by dropping the gauge shift on the polarization vectors, $\hat{e}\rightarrow e$, but manually imposing $e_i.p_j=e_j.p_i=0$ after performing the shift. The shifts are then equivalent, and in the rest of this paper the second shorter version will be used. It is also worth noting that the shifts are gauge invariant in $i$ and $j$, but non-local due to the extra poles.

It can be checked, though we do not prove, that any gluon amplitude satisfies the usual BCFW behavior under this shift. That is, under any shift $[i,j\rangle$:
\begin{align}
\begin{split}
\label{beha}
 A_n&\propto\mathcal{O}(z^{-1}) \textrm{ for $i$ and $j$ adjacent} \, ,\\
 A_n&\propto\mathcal{O}(z^{-2})  \textrm{ for $i$ and $j$ non-adjacent} \, .
\end{split}
\end{align}
In this case there is no ``bad shift", common to the on-shell method, as that is merely a by-product of an asymmetry imposed on the $\lambda$'s and $\tilde{\lambda}$'s. This shift can be used in the usual way to build general gluon amplitudes. For example, with a $[1,n\rangle$ shift:
\begin{equation}
A_n=\sum \frac{A^L_{i+1}(\hat{1},2,...,i, P)A^R_{n-i+1}(-P,i+1,...,n-1,\hat{n})}{P_i^2}\, ,
\end{equation}
where $z_i=\frac{P^2_i}{\hat{e}_i. P_i}$, and the usual summing over internal polarizations can be done using $\sum e^\mu e^\nu =\eta^{\mu\nu}$. Starting from the three point amplitude (which is just the three point Feynman vertex):
\begin{align}\label{3pt}
A_3(1,2,3)=V_3(1,2,3)=e_1.e_2 e_3.p_1+e_2.e_3 e_1.p_2-e_3.e_1 e_2.p_1\, ,
\end{align}
the four point amplitude can be obtained from a $[1,4\rangle$ shift as:
\begin{equation}\label{4bcfw}
A_4(1,2,3,4)= \frac{A_3(\hat{1},2,P)A_3(-P,3,\hat{4})}{p_1.p_2}\, ,
\end{equation}
with $z=p_1.p_2/\hat{e}_1.p_2$. It is easy to verify this is equal to the known amplitude, which in terms of Feynman diagrams is given by:
\begin{align}\label{feyn}
A_4(1,2,3,4)=\frac{V_3(1,2,P)V_3(-P,3,4)}{p_1.p_2}+\frac{V_3(1,4,P)V_3(-P,3,2)}{p_1.p_4}+V_4(1,2,3,4)\, ,
\end{align}
where $V_4=e_1.e_3\, e_2.e_4$. Comparing eqs. (\ref{4bcfw}) and (\ref{feyn}) makes clear the purpose of the non-local pole contained in the shifts: it generates the $p_1.p_4$ pole of the other channel. The computational advantage of this approach comes from the fact that fewer BCFW terms have to be considered compared to Feynman diagrams. In general, to compute an $n$ point amplitude there will be just $n-3$ BCFW terms to write down, compared to the factorially growing number of Feynman diagrams.

\section{Uniqueness from BCFW and locality}\label{sec3}
Besides the usual application of this shift to recursion relations, we conjecture that in fact the Yang-Mills scattering amplitude is the unique local object of mass dimension $[4-n]$ compatible with the usual BCFW behavior (\ref{beha}). As in Ref. \cite{Nima}, we start with an ansatz of local functions:
\begin{align}
M_n(p^{n-2})=\sum_i \frac{N_i(p^{n-2})}{\prod_{\alpha_i}P_{\alpha_i}^2}\, ,
\end{align}
where the sum is taken over all ordered cubic diagrams $i$, and $\alpha_i$ correspond to the channels of diagram $i$. The numerators $N_i$ are general polynomials of mass dimension $[n-2]$, and are linear in $n$ polarization vectors, but carry no information of factorization. Then, by requiring vanishing at infinity in a sufficient number of shifts, we obtain a unique solution, the gluon amplitude $A_n$. Empirically, it turns out that some shifts can be ignored completely, and the amplitude is still fixed. For example, at four points three shifts (for example, $[1,2\rangle$, $[2,1\rangle$, and $[2,3\rangle$) are enough to fix the answer, while at five points five shifts are needed. Furthermore, the required behavior in some shifts can be relaxed, and still the amplitude is fixed.

For the purposes of this proof, we will impose the maximal number of shifts, that is for all pairs $i$ and $j$ from $1$ to $n$, but with one crucial modification. For some shifts we will impose weaker constraints: under all the shifts involving some particle $h$, we will demand only $\mathcal{O}(z^0)$ for adjacent, and $\mathcal{O}(z^{-1})$ for non-adjacent shifts. This modification will be necessary for the inductive argument, which is carried out precisely by taking the special particle $h$ soft. We leave to future work the issue of finding the minimal set of shifts which fixes the amplitude.

\subsection{Overview of the proof}
It will be useful to introduce the following notation. Let $E_n$ be the set of all polarization vectors at $n$ points, and call $\mathcal{G}_n^h(E_n)$ the constraints (\ref{beha}), relaxed for particle $h$. Then we would like to prove that if $\mathcal{G}_n^h(E_n)$ for all $h=\overline{1,n}$ uniquely fixes $A_{n}(E_n,p^{n-2})$, the equivalent higher point set $\mathcal{G}_{n+1}^{h'}(E_{n+1})$ uniquely fixes $A_{n+1}(E_{n+1},p^{n-1})$, for all $h'=\overline{1,n+1}$. We will prove this statement for the choice $h'=n+1$, under the assumption that $A_n$ is fixed by both $\mathcal{G}_n^1(E_n)$ and $\mathcal{G}_n^n(E_n)$. All other choices for $h'$ can be treated in an identical manner, by taking $h'$ soft.

The basic logic of the argument is identical to that in Ref. \cite{Nima}. We consider a general local object at $n+1$ points, $M_{n+1}(p^{n-1})\delta_{n+1}$, where $\delta_{n+1}\equiv \delta(\sum_i p_i)$ is the $n+1$ point momentum conserving delta function, and show that imposing our constraints forces $M_{n+1}=A_{n+1}$, order by order in the soft expansion. Let $e_{n+1}=e$, $p_{n+1}=z\, q $, and expand $M_{n+1}\delta_{n+1}$ around $z=0$. Using momentum conservation to express $p_3$ in terms of the other momenta, the leading $1/z$ term has the general form:
\begin{align}
\label{big}
M^{-1}_{n+1}=\frac{\sum_{i} e.e_i B^i_{n;1}+\sum_{i\neq 3}e.p_i C^i_{n;1}}{q.p_1} +\frac{\sum_i e.e_i B^i_{n;n}+\sum_{i\neq 3}e.p_i C^i_{n;n}}{q.p_n}\, ,
\end{align}
where $B_{n;h}^i=B_{n;h}^i(\{e_1,e_2,\ldots e_n\}\setminus e_i,p^{n-1})$ and $C_{n;h}^i=C_{n;h}^i(\{e_1,e_2,\ldots e_n\},p^{n-2})$, with $h=1,n$, are local functions at $n$-points. We will show that imposing the BCFW constraints (\ref{beha}), relaxed for particle $n+1$, uniquely fixes:
\begin{align}
\label{leading}
M^{-1}_{n+1}= \left(\frac{e.p_1}{q.p_1}-\frac{e.p_n}{q.p_n}\right)A_{n}\, ,
\end{align}
which is the known leading piece of the Yang-Mills scattering amplitude \cite{Weinberg}. In principle, the argument would then continue to show that all subleading order terms of the object $M'_{n+1}\equiv M_{n+1}-A_{n+1}$ vanish, impying that $M_{n+1}=A_{n+1}$, completing the induction.

Formula (\ref{big}) also reveals why we must carry around this extra modification due to $h$. Shifts involving $h$ also shift the $q.p_h$ poles, producing one power of $z$ in the denominator. For example, if $M_{n+1}\propto z^{-1}$ under a shift $[1,2\rangle$, we should expect $B_{n,1}\propto z^{0}$ and  $C_{n,1}\propto z^{0}$ (ignoring the prefactors). But then in order for the inductive argument to close, the $n+1$ point constraints must also include such modified behaviors under particular shifts, which should not carry over to the lower point functions. Otherwise such relaxations would keep accumulating whenever we take a soft limit. This is precisely why particle $h$ is the one taken soft. After a soft limit $p_{n+1}\rightarrow 0$ we cannot not impose any shifts involving particle $n+1$, as it would not be consistent with momentum conservation. This is the only way to ensure that all functions always have just one particle  with relaxed constraints.

The proof will have four steps:
\begin{enumerate}
\item We show by induction that all functions $B^i_{n;h}$ in (\ref{big}) are ruled out.
\item We show that the $C^i_{n;h}$ are fixed by the uniqueness assumption at $n$-points, such that $C_{n;1}^i=a_i A_n$, and $C_{n;n}^i=b_i A_n$.
\item Using shifts involving particle $3$, which was chosen to impose momentum conservation, we show that $a_i=0$ for $i\neq 1$, and $b_i=0$ for $i\neq n$.
\item Finally, we use the $[1,n\rangle$ shift to fix $a_1=-b_n$. This shift is special because it is adjacent in $A_n$, but non-adjacent in $A_{n+1}$.
\end{enumerate}

The first step is the most laborious, and is carried out in section (\ref{B}). The last three steps are completed in section (\ref{C}).

\subsection{Ruling out $B(E_n^a)$ functions}\label{B}

First, to prove that the $B^i_{n;h}(E_n^{n-1},p^{n-1})$ functions are ruled out, we will have to consider the whole set of functions:
\begin{align}
\label{type}
B_{n;h}(a)\equiv B_{n;h}(E_n^a,p^{2n-2-a})\delta_n,\ a=\overline{0,n-1}\, ,
\end{align}
linear in just $a$ polarization vectors, which form the set $E_n^a$. The second lower index designates precisely the special particle $h$ mentioned above. In this case we can have $h=1$ or $h=n$.

Now that $E_n^a$ does not contain all polarization vectors at $n$ points, we would like to find general constraints $G_{n}^h(a)$ which rule out a function $B_{n;h}(a)$, and that also induct correctly. That is, $G_{n+1}$ constraints imposed on $B_{n+1}$ should imply $G_n$ constraints imposed on $B_n$. Furthermore, for $a=n-1$ the constraints should become the set we obtain by imposing $G_{n+1}(n)$ on $M_{n+1}$ in eq (\ref{big}).

The expected BCFW behavior will change based on what polarization vectors are missing. Let $\overline{i}$ designate particle $i$ if $B_{n;h}(a)$ is not a function of $e_i$, ie $e_i\notin E_n^a$. We will show that $B_{n;h}(a)$ functions cannot satisfy the following general $\mathcal{G}_n^h(a)$ constraints:
\begin{itemize}
\item $i$ and $j$ adjacent:
\begin{subequations}
\begin{itemize}
\item $[i,j\rangle,[j,i\rangle\propto z^{-1}$\, , \eqnum\label{c1}
\item $[i,\overline{j}\rangle\propto z^{-1}$\, ,\eqnum\label{c2}
\item $[\overline{i},j\rangle\propto z^{1}$\, , \eqnum\label{c3}
\item $[\overline{i},\overline{j}\rangle\propto z^{0}$ and $[\overline{j},\overline{i}\rangle\propto z^{1}$\, , \eqnum\label{c4}
\end{itemize}
\end{subequations}
\item $i$ and $j$ non-adjacent: a $z^m$ from above becomes $z^{m-1}$\, , \eqnum\label{c5}
\item shifts containing particle $h$: a $z^m$ from above becomes $z^{m+1}$\, . \eqnum\label{c6}
\end{itemize}
The cases (\ref{c1})-(\ref{c4}) pertain to the polarization structure, while the modifications (\ref{c5}) and (\ref{c6}) are related to the pole structure.

It can be verified explicitely that at four points these constraints rule out all functions:
\begin{align}
\begin{split}
\label{4p}
&a=0:B_{4,h}(p^6)\, ,\\
&a=1:B_{4;h}(E_4^1,p^5)\, ,\\
&a=2:B_{4;h}(E_4^2,p^4)\, ,\\
&a=3:B_{4;h}(E_4^3,p^3)\, ,
\end{split}
\end{align}
with $h=1$ and $h=4$. To be specific, $a=1$ functions include $B(e_1,p^5)$, $B(e_2,p^5)$ and so on, while functions with $a=2$ include $B(e_1,e_2,p^4)$, $B(e_1,e_3,p^4)$, and so on.
Now we move to the inductive step, and assume that $B_{n}$ functions (\ref{type}) are indeed ruled out by the $n$-point constraints. Then we must show this implies the higher point $B_{n+1}$ functions are ruled out by the $(n+1)$-point constraints. The $(n+1)$-point versions of the functions (\ref{type}) have the form:
\begin{align}
\label{npoint}
B_{n+1}(E_{n+1}^{a},p^{2n-a})\delta_{n+1}, \ a=\overline{0,n}\, .
\end{align}
However, a function $B_{n+1}$ is not necessarily a function of $e_{n+1}$, just like not all functions (\ref{4p}) contain $e_4$. The absence of $e_{n+1}$ changes the form of the soft limit, so we must treat each case separately.

\subsubsection{ Functions with $e_{n+1}\in E_{n+1}^a$}If $e_{n+1}\in E^a_{n+1}$, the functions (\ref{npoint}) can be written as:
\begin{align}
\label{orin}
B_{n+1}(a)\equiv B_{n+1}(\{E_{n}^{a},e_{n+1}\},p^{2n-a-1})\delta_{n+1},\ a=\overline{0,n-1}\, .
\end{align}
Again let $e_{n+1}=e$. The soft expansion of a function (\ref{orin}) has the same form of (\ref{big}):
\begin{align}
\label{short}
B_{n+1}(a) \rightarrow\sum_{h=1,n} \frac{1}{q.p_h}\left(\sum_r e.e_r B^{r}_{h}(a-1)+\sum_r e.p_r C^r_{h}(a)\right)\, .
\end{align}
Therefore a function $B_{n+1}(a)$ vanishes only if all functions $B_{h}(a)$ and $B_{h}(a-1)$ also vanish, explaining why we needed to consider the whole tower of functions in (\ref{type}).

It is easy to see that, because of the different denominators, functions in one pole do not mix with functions in the other pole under any shifts except $[1,n\rangle$ and $[n,1\rangle$, so we can treat the numerators as independent. The $[1,n\rangle$ and $[n,1\rangle$ shifts will be discussed separately.

We want to show that $G_{n+1}^{n+1}(a)$ constraints on $B_{n+1}(a)$ imply $G_{n}^h(a)$ constraints on $B_{n;h}(a)$, or in other words that any shift $[i,j\rangle$ inducts appropriately. Consider a function $B_n^k$ corresponding to the $q.p_n$ pole in eq. (\ref{short}), with $i,j\neq k$. Then we can write out the numerator of $q.p_n$ pole term:
\begin{align}\label{short2}
\nonumber &\sum_{r\neq i,j,k} e.e_r B_n^r+e.e_i B_n^i+e.e_jB_n^j+e.e_k B_n^k\\
&+\sum_{r\neq i,j,k} e.p_r C_n^r+e.p_i C_n^i+e.p_jC_n^j+e.p_k C_n^k\, .
\end{align}
Now assume that both $B_{n+1}$ and $B_{n}^k$ are functions of $e_i$ and $e_j$. Then this shift belongs to case (\ref{c1}) for both functions. We must show that if $B_{n+1}\propto z^{-1}$ under this shift, this implies that $B_n^k$ must have the same behavior. We can express this condition as:
\begin{align}
[i,j\rangle[B_{n+1}]\propto z^{-1}\Rightarrow [i,j\rangle[B_n^k]\propto z^{-1}\, .
\end{align}
To see this is the case, we apply the shift to eq. (\ref{short2}):
\begin{align}
\nonumber z^{-1}\propto &\sum_{r\neq i,j,k} e.e_r B_n^r+e.e_i B_n^i+(e.e_j+ z e.p_i\frac{e_i.p_j}{p_i.p_j})B_n^j+e.e_k B_n^k\\
&+\sum_{r\neq i,j,k} e.p_r C_n^r+(e.p_i+z e.e_i) C_n^i+(e.p_j- z e.e_i)C_n^j+e.p_k C_n^k\, .
\end{align}
The prefactor $e.e_k$ remains unique so $B_n^k$ cannot cancel against any of the other functions, and so must carry the same $z^{-1}$ behavior as $B_{n+1}$.

Similar reasoning can be applied for all the other cases. For some shifts however, such as $[k,i\rangle$, which is case (\ref{c1}) for $B_{n+1}$, but becomes a $[\overline{k},i\rangle$ case (\ref{c3}) for $B^k_n$, the argument will not be so simple: the shift mixes several functions together. The constraint can luckily be disentangled, and we can still obtain a (weaker, but necessary) constraint for $B_n^k$. This case is treated in appendix \ref{appa}, and all the others can be derived using identical arguments.

Next, we have to show that the modifications due to the pole structure also induct correctly. This is easy to see, since shifts involving $1$ or $n$ also shift the $q.p_1$ and $q.p_n$ poles in eq. (\ref{short}), contributing one power of $z$ in the denominator. Therefore these shifts weaken any constraints found above by one power of $z$. Finally, we have to show that the $[1,n\rangle$ shift also transforms accordingly. This is covered in appendix \ref{appb}.

Therefore all $B_{n;h}$ functions in (\ref{short}) vanish, and then the same reasoning can be applied for the $C_n^k$ functions, which will also vanish. This proves that all functions of the type $B_{n+1}(E_{n+1}^a)$ with $e_{n+1}\in E_{n+1}^a$ vanish.

\subsubsection{ Functions with $e_{n+1}\notin E_{n+1}^a$}In this case $E_{n+1}^a=E_n^a$ and we must consider all functions of the type:
\begin{align}
\label{notin}
B_{n+1}(a)=B_{n+1}(E_n^a,p^{2n-a}),\ a=\overline{0,n}\, .
\end{align}
In the absence of $e_{n+1}$, the soft limit is given by a simpler expression:
\begin{align}
\label{new}
B_{n+1}(a)\rightarrow \sum_{h=1,n}\frac{1}{q.p_h} B_{n;h}(E_n^a,p^{2n-a})\, ,
\end{align}
but now the functions on the right side are not of the type (\ref{type}): they have two extra powers of momenta. We can distinguish between functions with $a<n$ and $a=n$, which we denote:
\begin{align}
\label{bprim}B'(a)&\equiv B_{n;h}(E_n^a,p^{2n-a})\, ,\\
\label{bnprim}B'&\equiv B_{n;h}(E_n^n,p^{n})\, .
\end{align}

For $a<n$, the functions can be written in terms of the previous functions (\ref{type}), as $B'(a)=B(a)\sum a_{ij} p_i p_j$, since they have more momenta than polarization vectors. Then it is easy to show that if $B(a)$ functions are ruled out by $G_n^h(a)$ constraints, so must the $B'(a)$ functions. The details of this proof are given in appendix \ref{appc}.

Finally, the functions with $a=n$, $B'$, cannot be expressed in terms of $B_n(E_n^n,p^{n-2})$. However, the higher point version of such functions, $B_{n+1}(E_{n+1}^{n+1},p^{n+1})$, is always a function of $e_{n+1}$, so if we use the soft limit:
\begin{align}
B_{n+1}(E_{n+1}^{n+1},p^{n+1})\rightarrow \sum_{h=1,n}\frac{1}{q.p_h}\left(\sum e.e_r B_h^r(E_n^{n-1},p^{n+1})+e.p_r C_h^r(E_n^n,p^n)\right)\, ,
\end{align}
we are guaranteed to land only on functions which were already shown to vanish. The $B$ functions are of type (\ref{bprim}) with $a=n-1$ and were shown to vanish, while the $C$ functions are just the original $n$-point functions (\ref{bnprim}), and vanish by assumption. Therefore $B_{n+1}$ functions (\ref{notin}) with $e_{n+1}\notin E^a_{n+1}$ also vanish (at the leading level) under the corresponding constraints.

In conclusion, all possible types of $B_{n+1}$ functions vanish under $G_{n+1}$, concluding the inductive proof that all functions $B_{n;h}(E_n^a,p^{2n-a-2})$ vanish under $\mathcal{G}_n^h(E_n^a)$ constraints, including those in our original eq. (\ref{big}).

\subsection{Fixing $C(E_n^n)$ functions }\label{C}
Once the $B$ functions in (\ref{big}) have been shown to vanish, using the same arguments as before it is easy to see that the $C$ functions must satisfy $\mathcal{G}_n^h(E_n^n)$ constraints. But by assumption functions of the type $C_{n;h}(E_n^n,p^{n-2})$ are uniquely fixed by these constraints, up to some numerical coefficient. Therefore we obtain $C_1^i=a_i A_n$ and $C_n^i=b_iA_n$, and eq. (\ref{big}) becomes:
\begin{align}
M_{n+1}^{-1}=\sum_{i\neq 3}\left(a_i\frac{ e.p_i }{q.p_1} +b_i\frac{ e.p_i }{q.p_n}\right)A_n\, .
\end{align}

Now we exploit the fact that due to our choice of imposing momentum conservation, $p_3$ is not present in the sums above. Consider the $q.p_1$ pole term first. Under a shift $[i,3\rangle$, $i\neq 1$, only the prefactor $e.p_i$ is affected, $e.p_i\rightarrow e.p_i-z e.e_i$. Both $M_{n+1}$ and $A_n$ have the same scaling $\mathcal{O}(z^m)$ (with $m=-1$ for adjacent and $m=-2$ for non-adjacent), and we obtain:
\begin{align}
\mathcal{O}(z^m)\propto M_{n+1}^{-1}=& \frac{\sum_{j} a_j e.p_j }{q.p_1}\times A_n\propto \frac{\sum_{j} a_j e.p_j-z a_i e.e_i}{q.p_1}\times\mathcal{O}(z^m)=\mathcal{O}(z^m)+ a_i\mathcal{O}(z^{m+1})\, ,
\end{align}
which implies $a_i=0$, for $i\neq 1$. The coefficient $a_1$ is not ruled out by the same trick, since under $[1,3\rangle$ the pole is also shifted:
\begin{align}
\mathcal{O}(z^m)\propto M^{-1}_{n+1}= a_1\frac{e.p_1}{q.p_1} \times A_n\propto a_i \frac{e.p_1+z e.e_1}{q.p_1+z q.e_1} \times\mathcal{O}(z^{m})\propto a_1 \mathcal{O}(z^m)\, .
\end{align}
Similarly we obtain that $b_i=0$ for $i\neq n$, and we are left with:
\begin{align}
M^{-1}_{n+1}= \left( a_1 \frac{e.p_1}{q.p_1}+b_n \frac{e.p_n}{q.p_n}\right)A_n\, .
\end{align}

Finally, under the special $[1,n\rangle$ shift, which crucially is adjacent in $A_{n}$ but non-adjacent in $M_{n+1}$:
\begin{align}
\nonumber \mathcal{O}(z^{-2})\propto M^{-1}_{n+1}=\left( a_1 \frac{e.p_1}{q.p_1}+b_n \frac{e.p_n}{q.p_n}\right)\times A_n&\propto \left((a_1+b_n)\frac{e.e_1}{q.e_1}+\mathcal{O}(z^{-1})\right)\times \mathcal{O}(z^{-1})\\
&\propto (a_1+b_n)\mathcal{O}(z^{-1})+\mathcal{O}(z^{-2})\, ,
\end{align}
so $a_1=-b_n$, and we obtain Eq. (\ref{leading}):
\begin{align}
M^{-1}_{n+1}= \left( \frac{e.p_1}{q.p_1}- \frac{e.p_n}{q.p_n}\right)A_n\, ,
\end{align}
completing the leading order proof.

\section{Future directions}
In this article we have provided the leading order step in the proof that locality and correct behavior under BCFW shifts uniquely fix the Yang-Mills tree-level amplitude. It is very likely that the subleading terms can be treated in the same way, but finding a more direct proof would be far more rewarding. The most direct approach would be to show that large $z$ BCFW shifts are somehow related to gauge invariance. Then the proof in Ref. \cite{Nima} would immediately apply. But this connection would be very surprising in its own right. For instance, it might help explain why Yang-Mills and gravity have this surprising behavior in the first place. One option in this direction would be to use the Cauchy theorem to build the amplitude from different shifts (but without assuming unitarity). It is very peculiar that these shifts are gauge invariant, but non-local, so the object they construct is guaranteed to unfortunately inherit both properties, and thus avoid the locality+gauge invariance argument. Yet somehow, constructing the object from many different shifts must eliminate the non-local terms. Ultimately, this result suggests the notion of ``constructability'' \cite{onshell}-\cite{He} might play a more fundamental role, beyond recursion relations.

Another obvious direction is determining whether an equivalent statement holds for gravity. In this case it is likely that demanding the stronger $\mathcal{O}(z^{-2})$ behavior will be required. We suspect this to be the case due to the fact that even though $\mathcal{O}(z^{-1})$ is sufficient for recursion relations to exist, it was discovered that the so called ``bonus'' behavior of gravity is actually required for full on-shell consistency \cite{onshell}. In Ref. \cite{bonus} it was shown that this bonus behavior automatically emerges from Bose-symmetry, but it was unclear whether the logic can be reversed: could $\mathcal{O}(z^{-2})$ behavior imply Bose-symmetry? If the claim of this present article can indeed be extended to gravity, then clearly the answer is yes. And not only would the Bose-symmetry emerge, but (assuming locality) the whole amplitude emerges.

Scalar theories  like the non-linear sigma model or Dirac-Born-Infeld are another obvious target. Recently, it was shown that recursion relations can be applied to such theories as well, when the soft behavior is taken into account \cite{nlsm}. On the other hand, in Ref. \cite{Nima} it was shown that locality and the soft behavior completely fix the amplitude. It would be very interesting to fully work out the interplay between locality, vanishing in the soft limit, and BCFW shifts, in the context of this article.

As a by-product of this investigation, we have also uncovered a new method for computing the full Feynman amplitude via BCFW recursion relations, a method which to our knowledge has not yet been explored. It would be interesting to see what applications might be derived from this new approach.

\section*{Acknowledgments}
The author would like to thank Nima Arkani-Hamed for suggesting an exploration of this BCFW shift, Jaroslav Trnka for comments, and Savan Kharel for pointing out a wrong sign in eq. (\ref{shift}).

\appendix

\section{Inducting the $[k,i\rangle$ shift} \label{appa}
In this section we show that the a shift $[k,i\rangle$ corresponding to case (\ref{c1}), imposes case (\ref{c3}) constraints on $B_n^k$, since $B_n^k$ is not a function of $e_k$. Succinctly, we want to show:
\begin{align}
\label{gg3}
[k,i\rangle[ B_{n+1}]\propto z^{-1}\Rightarrow [\overline{k},i\rangle[ B_n^k]\propto z^{1}
\end{align}
The numerator becomes (\ref{short}):
\begin{align}
\nonumber \mathcal{O}(z^{-1})\propto&\sum_{r\neq i,k} e.e_r B_n^r+(e.e_i+ z e.p_k\frac{e_k.p_i}{p_i.p_k})B_n^i +e.e_k B_n^k\\
&+\sum_{r\neq i,k} e.p_r C^r_n+(e.p_i- z e.e_k)C^i_n+(e.p_k+z e.e_k) C^k_n
\end{align}
in this case $e.e_k$ is no longer unique, so our constraint now involves other functions:
\begin{align}
\label{oo}
e.e_k(B_n^k+z(C_n^k-C_n^i))\propto\mathcal{O}( z^{-1})
\end{align}
However we can still obtain an upper bound for $B_n^k$, if we find one for $C_n^k$ and $C_n^i$.

Observe that under this shift, the prefactors $e.e_i$ and $e.p_i$ are unique, so $B^i,C^i\propto z^{-1}$. Next the term proportional to $e.p_k$ is:
\begin{align}
\label{fix}
e.p_k(C_n^k+z \frac{e_k.p_i}{p_i.p_k}B_n^i)\propto e.p_k(C_n^k+\mathcal{O}(z^{0}))\propto \mathcal{O}( z^{-1})
\end{align}
which implies $C_n^k$ can be at most $\propto \mathcal{O}( z^{0})$. Therefore eq. (\ref{oo}) becomes:
\begin{align}
B_n^k+ z(C_n^k-C_n^i)\propto B_n^k+ \mathcal{O}(z^{1})-\mathcal{O}(z^{0})\propto \mathcal{O}( z^{-1})
\end{align}
so $B_n^k$ is at most $\propto\mathcal{O}( z^{1})$ under this shift, as required.

\section{Inducting the $[1,n\rangle$ shift}{\label{appb}
In this section we consider the $[1,n\rangle$ and $[n,1\rangle$ shifts. These are non-adjacent in $B_{n+1}$, while in $B_n^k$ they are adjacent, and are also affected by the pole shift. We will consider just the case where both $B_{n+1}$ and $B_n^k$ are functions of $e_1$ and $e_n$. Therefore if $B_{n+1}\propto z^{-2}$ under the non-adjacent $[1,n\rangle$ shift, we expect to obtain $B^k\propto z^{0}$: one power from the shift in the denominator, and one from becoming an adjacent shift. So we want to show:
\begin{align}
[1,n\rangle[B_{n+1}]\propto z^{-2}\Rightarrow [i,j\rangle[B_n^k]\propto z^{0}
\end{align}
Since under this shift $B_n^k$ can cancel against $B_1^k$, we can no longer treat the numerators as independent, so we must consider both functions in (\ref{short}):
\begin{align}
z^{-2} \propto [1,n\rangle\left[\frac{e.e_k B_1^k}{q.p_1}+\frac{e.e_k B^k_n}{q.p_n}\right]&= e.e_k\left(\frac{ B^k_1}{q.p_1+z q.e_1}+\frac{ B^k_n}{q.p_n-zq.e_1}\right)=\\
&=\frac{1}{z} \frac{B^k_1-B^k_n}{q.e_1}-\frac{1}{z^2}\frac{q.p_1B^k_1-q.p_n B^k_n}{(q.e_1)^2}+\mathcal{O}(z^{-3})
\end{align}
which implies $\frac{1}{z^{2}} \frac{q.p_nB^k_n}{(q.e_1)^2}\propto z^{-2}$ due to the unique prefactor, so $B_n^k\propto z^{0}$, which is the expected result.

\section{Ruling out $B'(a)$ functions with extra momenta}\label{appc}
We will show that $B'(a)$ functions (\ref{bprim}) vanish under $G_n^h(a)$ constraints if $B(a)$ functions (\ref{type}) also vanish. We can express the former in terms of the latter as:
\begin{align}
\label{extra}
B'(a)=B(a)\sum_{r,s} a_{rs}p_r.p_s
\end{align}
Then we must show:
\begin{align}
\label{precise}
[i,j\rangle[B'(a)]\propto z^m \Rightarrow [i,j\rangle[B(a)]\propto z^m
\end{align}
Under a shift $[i,j\rangle$, with $r,s\neq i,j$, we have:
\begin{align}
\begin{split}
p_r.p_s &\rightarrow p_r.p_s\\
p_i.p_j &\rightarrow p_i.p_j\\
p_i.p_r&\rightarrow p_i.p_r-z e_i.p_k\\
p_j.p_r&\rightarrow p_j.p_r+z e_i.p_k
\end{split}
\end{align}
Assume that under this shift $B'(a)\propto \mathcal{O}(z^m)$, but that $B(a)$ goes like a higher power, $\mathcal{O}(z^{m+1})$. Equation (\ref{extra}) becomes:
\begin{align}
\nonumber z^m B'^0+\mathcal{O}(z^{m-1})=&\left(z^{m+1}B^1+z^m B^0+\mathcal{O}(z^{m-1})\right)\left(z \sum_r (a_{jr}-a_{ir})e_i.p_r+\sum_{rs}a_{rs}p_r.p_s\right)=\\
\nonumber=&z^{m+2}\left(B^1 \sum_r e_i.p_r (a_{jr}-a_{ir})\right)\\
\nonumber&+z^{m+1}\left(B^1\sum_{rs}a_{rs}p_r.p_s+ B^0 \sum_r (a_{jr}-a_{ir})e_i.p_r\right)\\&+\mathcal{O}(z^m)
\end{align}
On the left side we have only $\mathcal{O}(z^m)$, so the higher orders on the right must vanish. Order $z^{m+2}$ vanishing implies either $B^1=0$, or $\sum_r e_i.p_r (a_{jr}-a_{ir})=0$. In the latter case, vanishing at order $z^{m+1}$ implies $B^1=0$. Either way then we must have $B\propto z^m$, proving eq. (\ref{precise}), and so functions of the type (\ref{bprim}) are ruled out.

\end{document}